# MESH RESOLUTION EFFECT ON 3D RANS TURBOMACHINERY FLOW SIMULATIONS

SERGIY YERSHOV [1,*], VIKTOR YAKOVLEV [2]

[1] *self-employed researcher, Savoy, IL, USA*
[2] *Institute for Mechanical Engineering Problems of NASU, Kharkiv, Ukraine*

**Abstract.** The paper presents the study of the effect of a mesh refinement on numerical results of 3D RANS computations of turbomachinery flows. The CFD solver F, which based on the second-order accurate ENO scheme, is used in this study. The simplified multigrid algorithm and local time stepping permit decreasing computational time. The flow computations are performed for a number of turbine and compressor cascades and stages. In all flow cases, the successively refined meshes of H-type with an approximate orthogonalization near the solid walls were generated. The results obtained are compared in order to estimate their both mesh convergence and ability to resolve the transonic flow pattern. It is concluded that for thorough studying the fine phenomena of the 3D turbomachinery flows, it makes sense to use the computational meshes with the number of cells from several millions up to several hundred millions per a single turbomachinery blade channel, while for industrial computations, a mesh of about or less than one million cells per a single turbomachinery blade channel could be sufficient under certain conditions.

## 1. Introduction

Currently, the computational fluid dynamics (CFD) is an essential tool for investigating turbomachinery flows and designing turbine and compressor flowpathes. The most common current approach is a numerical simulation of turbulent viscous compressible flows using the Reynolds-averaged Navier-Stokes (RANS) equations. It is considered that such a problem statement, as well as the used numerical methods and algorithms are sufficiently mature whereas the corresponding CFD solvers are thoroughly tested and adjusted [1].

Nevertheless, accuracy and reproducibility of the numerical results often leave much to be desired. The numerical results of the turbomachinery flow simulations can be determined by the computational mesh characteristics. A quantitative and qualitative accuracy of the numerical solutions depends on a cell shape, size of the nearest cell to the wall $y^+$, the number of cells across boundary layer, the mesh expansion ratio, the curvature and non-smoothness of mesh lines, etc.

Over the past twenty years, guidelines for choosing the mesh resolution for the numerical simulation of turbomachinery viscous flows using the RANS models have changed several times: from 100-200 thousand cells per one blade channel in 90-ies of the last century up to 0.5-1.0 million cells per one blade channel now [1–3]. Usually, a mathematical basis of such recommendations is not clear, requirements to the mesh refinement are often not well-founded [3] (perhaps, with the only exception for $y^+$) and the question of the solution convergence remains open.

The present paper continues the previous authors' study [4] and considers issues of a choice of the mesh refinement and its influence on results of the 3D viscous turbomachinery flow computations.

## 2. Formulation of research problems

We have performed the RANS simulation of the 3D turbulent compressible viscous flow through several turbomachinery stages and cascades. The $k$–$\omega$ SST turbulence model [5] was used in this work. The main objective was a qualitative study of the numerical solution convergence without being tied to the experimental data. It is evident that both the insufficient adequacy of the mathematical model as well as approximation and experimental errors could lead to the fact that in some cases the differences between the numerical results and the experimental data may increase as the mesh is refined.

---

[*] First Corresponding Author. Email: sergiy.v.yershov@gmail.com

The numerical simulation was performed using the CFD solver F [6, 7], which is based on the implicit second-order ENO scheme [8, 9] and a simplified multigrid algorithm. The local time step was used for the convergence acceleration. During the initial stages of the computations the CFL number was chosen in the range from 30 to 50, and just before the convergence achievement it was reduced to 5-10 in order to improve the solution accuracy. At that, during the whole computations the time step for excessively elongated cells was somewhat reduced in a special way [10].

The simplified multigrid algorithm (SMA) contains in using the set of the successively nested meshes (usually, 4 or 5 nesting levels) for each flow computation. At that, changing the current nesting level by one leads to the change of the number of cells in each direction strictly twice. The flow computations start on the coarsest of the successively nested meshes. As the convergence occurs the numerical results are interpolated to the next finer mesh. This procedure is repeated until the solution convergence on the finest mesh is reached. It should be noted that the solutions on the successively nested meshes of the SMA, can not be used for the analysis of the mesh convergence in the case of turbulent flow because ensuring the condition $y^+ \approx 1$ for all mesh levels simultaneously is impossible.

The constancy of the kinetic energy losses with a given accuracy was considered as the main criterion of the temporal convergence of the computations. Additionally, at the inlet and the exit of each blade row we controlled the convergence of both mass-flow-rate and the turbulent kinetic energy fluxes. When the constancy of the blade forces was used as the convergence criterion, errors in determining the cascade or stage efficiency, as it turned out, may be up to 0.01-0.02 (1-2 percentage points). It is also important to note that the solution convergence defects, made during the computations on the coarser meshes using SMA, are very slowly eliminated at the finer meshes, especially in small and highly elongated cells, which are generally located in the boundary layers and wakes. Slow changes of flow characteristics in the boundary layers usually have a little effect on the blade forces that can create an illusion of the temporal convergence of the solution. As a result, the kinetic energy losses estimate may be significantly wrong. Therefore, during the computations using SMA it is extremely important to ensure a careful convergence of the solution at all levels of the nested meshes.

We used the meshes of H-type with an approximate orthogonalization near solid walls. The meshes considered in this study were conventionally divided into five groups based on the number of cells per one blade channel:
1) very coarse meshes of $10^4$–$10^5$ cells;
2) coarse meshes of $10^5$–$10^6$ cells;
3) intermediate meshes of $10^6$–$10^7$ cells;
4) fine meshes of $10^7$–$10^8$ cells;
5) very fine meshes of $10^8$–$10^9$ cells.

During the present mesh convergence study the number of cells in each spatial direction increased about twice. For the meshes of all considered groups value of $y^+$ both in the radial and circumferential directions was set close to unity. It was found that an adequate prediction of the law of the wall (universal velocity profile) is possible only if at least 30 cells are placed across the boundary layer and the mesh expansion ratio in the wall-normal direction does not exceed 1.1. Therefore, when we performed computations on meshes of the groups 3, 4, and 5, these requirements were strictly enforced. Ensuring these requirements on meshes of the groups 1 and 2 without reducing the accuracy in the flow core is very problematic, so in these cases trade-off decisions were taken.

It should be noted that the computations using meshes of more than one million cells require considerable computational time and are almost impossible without the mechanisms of the convergence acceleration [9] implemented in the solver F.

Figure 1 presents an example of the meshes, considered in this study (it is shown only every fourth mesh line in each direction for clarity).

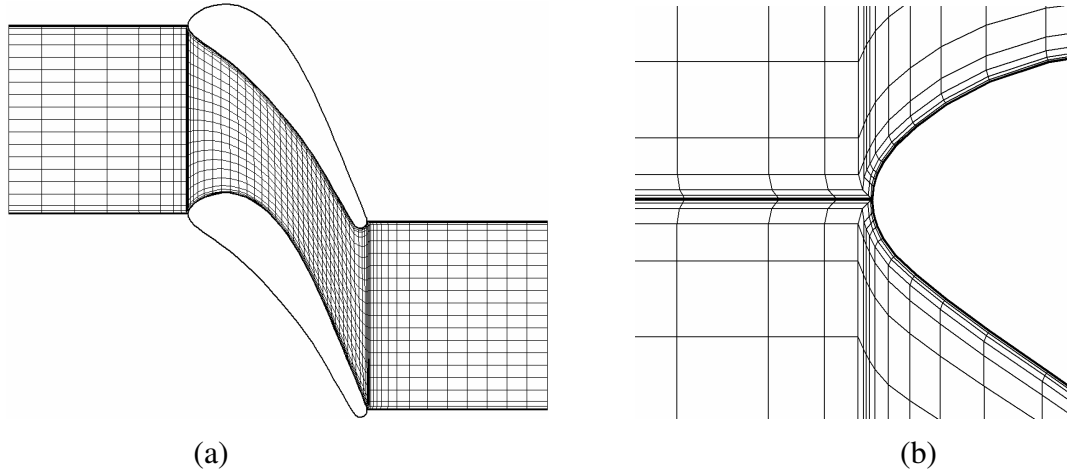

*Fig. 1. The computational mesh in the VKI-Genoa turbine cascade [11]*
*(only every fourth mesh line is shown)*
(a) tangential section; (b) mesh fragment near the leading edge

## 3. Subsonic flow within the VKI-Genoa cascade

We considered subsonic flow in the turbine cascade, that was experimentally investigated in Ref. [11]. The exit Mach number was $M_{1is} = 0.24$. We performed the computations using the meshes of 65 thousand cells, 524 thousand cells, and 4.2 million cells.

Figure 2 shows the Mach number contours at the midspan section in the cases of the meshes of 520 thousand cells and 4.2 million cells. It is clearly seen that the solutions are quite similar.

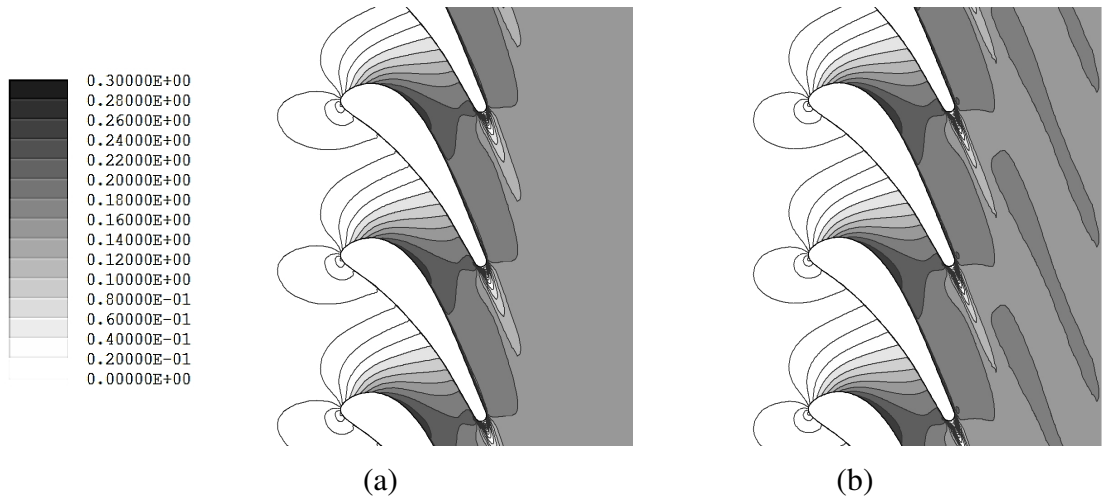

*Fig. 2. The Mach number contours at the midspan section of the VKI-Genoa turbine cascade*
(a) mesh of $5.2 \cdot 10^5$ cells; (b) mesh of $4.2 \cdot 10^6$ cells

Table 1 shows the values of the kinetic energy losses coefficient for the considered cascade. There is a trend of the solution convergence when the mesh is refined. It can be seen that the use of the mesh of the group 2 (Fig. 2,a) can lead to an error in the losses of about 0.005 (0.5 percentage points), compared with the numerical results obtained using the mesh of the 3$^{rd}$ group (Fig. 2,b).

Based on the above results, the authors have chosen the computational mesh for further investigating the laminar-turbulent transition in this cascade [12, 13].

The kinetic energy losses in the VKI-Genoa turbine cascade

| Mesh group | 1 | 2 | 3 |
|---|---|---|---|
| Number of cells | $6.5 \cdot 10^4$ | $5.2 \cdot 10^5$ | $4.2 \cdot 10^6$ |
| Kinetic energy losses | 0.188 | 0.127 | 0.123 |

## 4. Transonic flow through the ABB-Saturn turbine stage

We considered transonic flow within the turbine stage of the company ABB-Saturn. The geometry and flow conditions of the stage were kindly provided to one of the authors of this paper by Prof. A.V.Granovsky [14]. In this flow case the flow conditions were $M_{1is} = 0.87$ at the stator cascade exit and $M_{w2is} = 0.71$ at the rotor cascade exit. The computations were performed using the meshes of 30 thousand cells, 240 thousand cells, 1.9 million cells, and 15.3 million cells per each blade channel of the stator and rotor cascades.

Figures 3 and 4 show the numerical flow patterns at the midspan section of the stator and rotor cascades, respectively. The numerical Schlieren images are given for the different meshes, whereas the Mach number contours are presented only for the finest mesh.

It is clearly seen from the figures that in the case of the meshes of the group 2 (Figs. 3,a and 4,a) the shock waves are blurred and it is difficult to define their position precisely. In the case of the mesh of the 3$^{rd}$ group (Figs. 3,b and 4,b), the shocks are captured better. A clear shock wave pattern is obtained only using the mesh of the group 4 (Figs. 3,c and 4,c). The similar conclusions can be drawn for the resolution in the separation and wake regions.

It should be noted that here and further the numerical Schlieren images contain computational artifacts in the form of horizontal and vertical lines near the leading and trailing edges. This phenomenon takes place in the areas of significant changes in step and shape of cells and indicates the loss of accuracy, which manifests itself in a large error in determining derivatives of the flow parameters.

Table 2 shows the kinetic energy losses in the stage for the different meshes. There is a tendency of convergence when mesh is refined. Absolute errors of determining the kinetic energy losses in the case of the meshes of the groups 2 and 3 compared with the case of the mesh of the group 4 are 0.0014 (1.4 percentage points) and 0.005 (0.5 percentage points), respectively.

Table 2

The kinetic energy losses in the ABB-Saturn turbine stage

| Mesh group | 1 | 2 | 3 | 4 |
|---|---|---|---|---|
| Number of cells | $2 \times 3.0 \cdot 10^4$ | $2 \times 2.4 \cdot 10^5$ | $2 \times 1.9 \cdot 10^6$ | $2 \times 1.5 \cdot 10^7$ |
| Kinetic energy losses | 0.129 | 0.112 | 0.103 | 0.098 |

## 5. Supersonic flow within the compressor cascade of the EU FP7 research project TFAST

We considered supersonic flow through the compressor cascade, which was experimentally and computationally studied in the EU FP7 program project, TFAST [15]. According to the experimental conditions the boundary layer was extracted at the endwalls to reduce separations and secondary flows. The computations were carried out for the conditions $\pi_c = 1.22$ and $\pi_c = 1.87$ with the relative Mach number at the inlet $M_{1w} = 1.2$ using the meshes of 2.4 million cells, 9.2 million cells, 54 million cells, and 250 million cells. The first three meshes have the same number of cells and the mesh dimensions at the tangential section as the 2D meshes of Ref. [16].

Figure 5 demonstrates the numerical Schlieren images of the flow pattern at the midspan section in the case of the compression ratio $\pi_c = 1.22$ for the different meshes. It is seen that in the case of the meshes of the group 3 (Figs. 5,a and 5,b) all shock waves are very blurred. In the case of the mesh of the group 4 (Fig. 5,c) the resolution of the bow shock is improved, but the reflection of the leading-edge oblique shock impinging on the blade suction side as well as the non-smeared shock wave pattern near cascade throat and downstream are clearly captured only in the case of the mesh of the group 5 (Fig. 5,d). The mesh refinement improves significantly the separation and wake resolution, and especially capturing the tangential discontinuities downstream of the triple shock points. In the case of the mesh of the group 5 the numerical results are the best comparing with other considered results. A similar behavior of the numerical solution is observed in the case of the compression ratio $\pi_c = 1.87$. In the latter flow case, when using the meshes of the group 3, the error on the internal shock location can reach almost 0.1 blade chord.

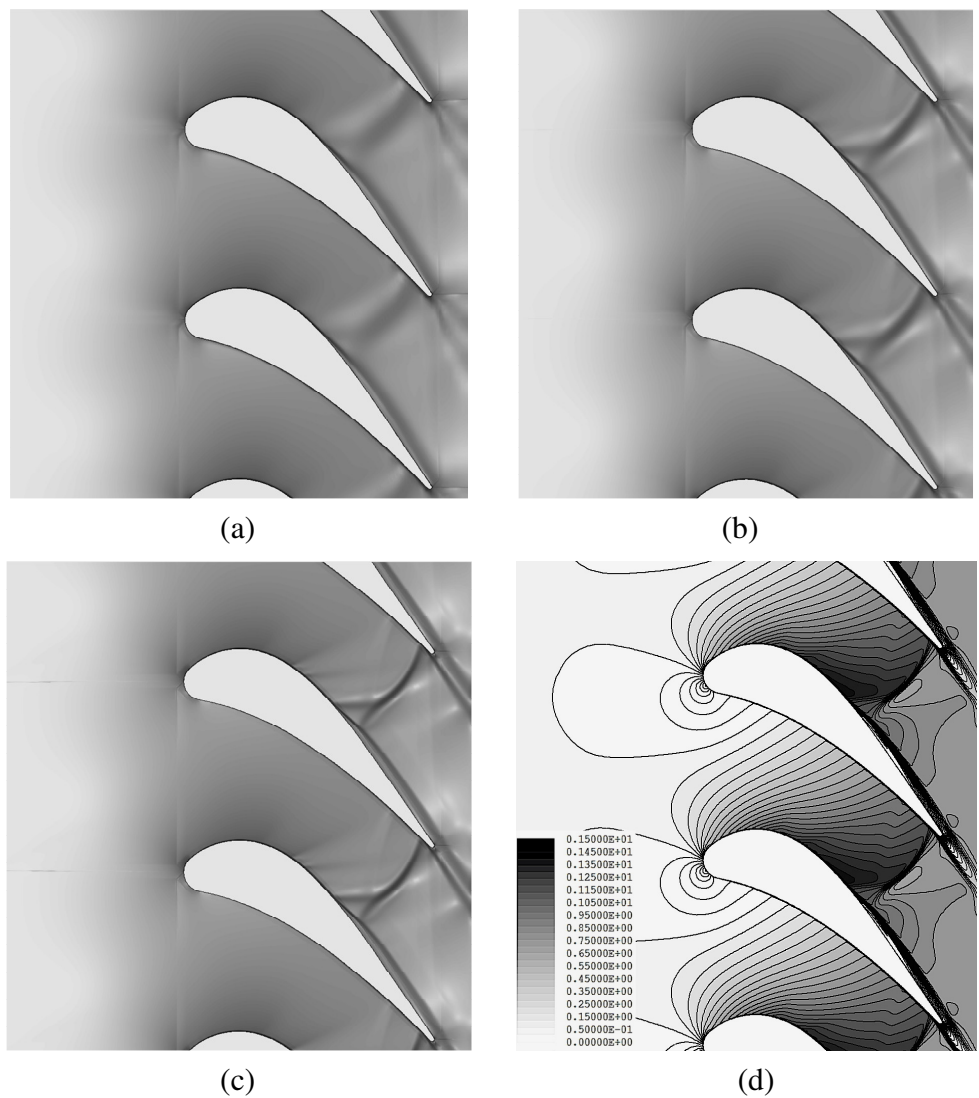

(a)            (b)

(c)            (d)

**Fig. 3. The flow pattern at the midspan section of the stator cascade of the ABB-Saturn turbine stage**
(a), (b), (c) numerical Schlieren (meshes of $2.4 \cdot 10^5$, $1.9 \cdot 10^6$ and $1.5 \cdot 10^7$ cells, respectively);
(d) Mach number contours (mesh of $1.5 \cdot 10^7$ cells)

Table 3 shows the kinetic energy losses in the cascade for the different meshes. There is a tendency of convergence when mesh is refined. It is seen that the absolute error in determining the kinetic energy losses in the case of the meshes of the group 3 compared with the case of the mesh of the group 5 could be more than 0.005 (0.5 percentage points).

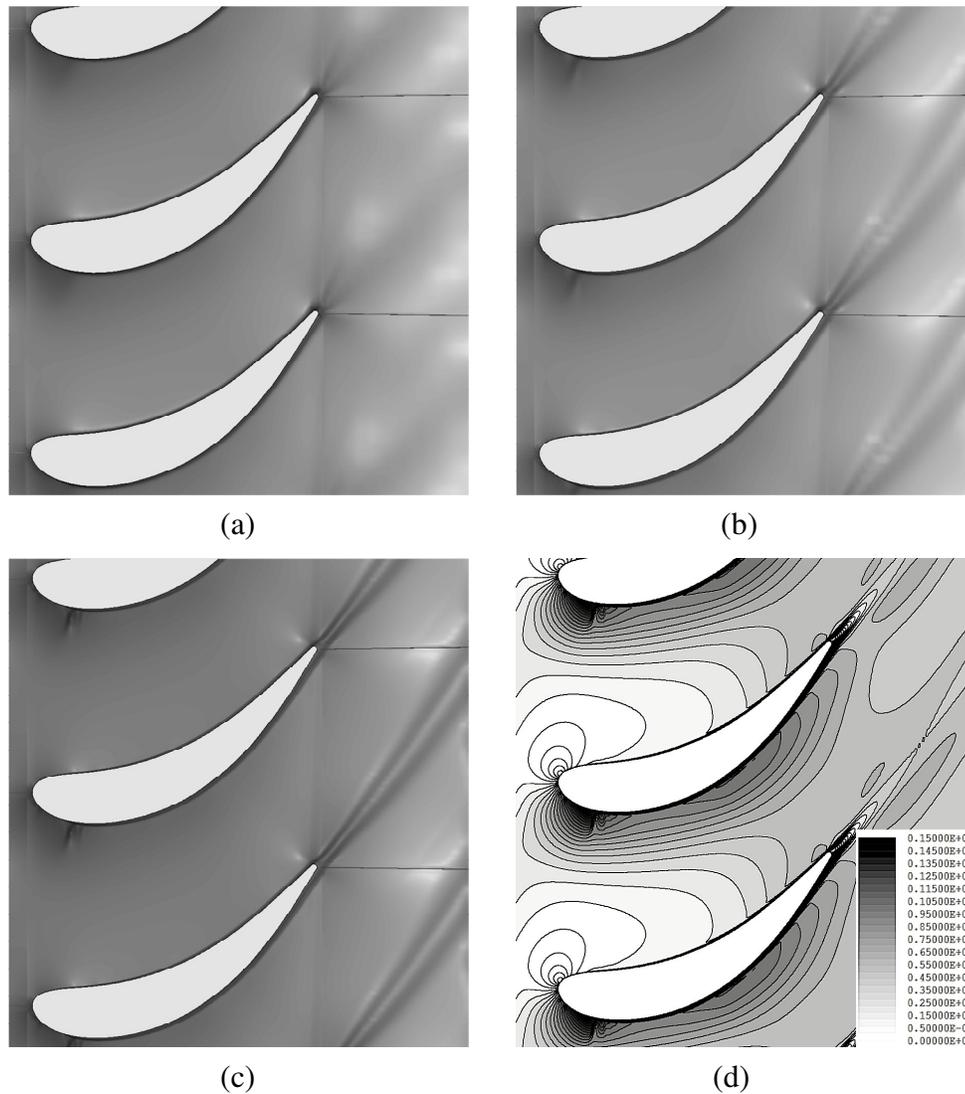

*Fig. 4. The flow pattern at the midspan section of the stator cascade of the ABB-Saturn turbine stage*
(a), (b), (c) – numerical Schlieren (meshes of $2.4 \cdot 10^5$, $1.9 \cdot 10^6$ and $1.5 \cdot 10^7$ cells, respectively);
(d) – Mach number contours (mesh of $1.5 \cdot 10^7$ cells)

Table 3
The kinetic energy losses in the TFAST compressor cascade for $\pi_c = 1.87$ conditions

| Mesh group | 3 | | 4 | 5 |
|---|---|---|---|---|
| Number of cells | $2.4 \cdot 10^6$ | $9.2 \cdot 10^6$ | $5.4 \cdot 10^7$ | $2.5 \cdot 10^8$ |
| Kinetic energy losses | 0.1396 | 0.1373 | 0.1342 | 0.1340 |

**6. Transonic flow within the turbine cascade of the EU FP7 research project TFAST**

We considered transonic flow in the turbine cascade, which was computationally and experimentally studied in the EU FP7 program project, TFAST [15]. In this flow case the exit Mach number was $M_{1is} = 1.05$. We calculated the cascade flow without and with film cooling using the meshes of 85 thousands cells, 680 thousands cells, 5.4 million cells, and 43.5 million cells. Film cooling was provided using air blowing through two rows of holes uniformly spaced along the blade (46 holes in each row). The area of each hole was about 0.5 mm$^2$, whereas the blade length and chord were 125 mm and 76 mm, respectively. In our computations, there were from 1 to 4 cells per one hole in the case of the mesh of the group 3 and from 6 to 16 cells per one hole in the case of the mesh of the group 4.

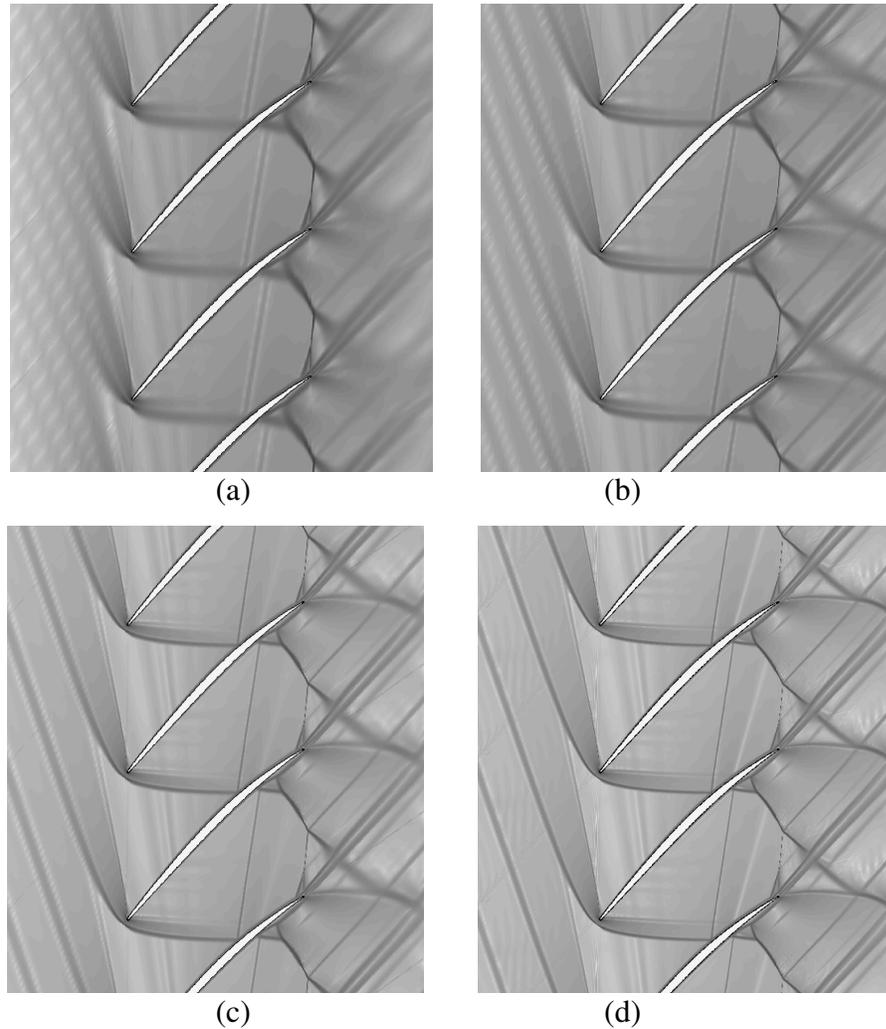

(a)          (b)

(c)          (d)

*Fig. 5. The numerical Schlieren images at the midspan section of the TFAST compressor cascade*
(a) $2.4 \cdot 10^6$ cells; (b) $9.2 \cdot 10^6$ cells; (c) $5.4 \cdot 10^7$ cells; (d) $2.5 \cdot 10^8$ cells

Figure 6 presents the Mach number contours at the midspan section of the cascade channel for the different meshes. In the case of film cooling, the image patches near the leading edges are only shown. These figures demonstrate that the mesh refinement improves significantly the resolution of the shock waves and wakes. So, in the case of the meshes of the groups 2 (Fig. 6,a) and 3 (Fig. 6,b) the resolution of the first shock wave at the suction side just downstream of the cascade throat is unsatisfactory, and only in the case of the mesh of the group 4 (Fig. 6,c) this shock wave can be definitely interpreted as a discontinuity.

The cooling gas jets are also better resolved on the finer meshes. We did not carry out the computations using the meshes of the groups 1 and 2, since in this case the faces of near-wall cells were larger than the holes. It is clearly seen from Fig. 6 that in the case of the mesh of the group 3

(Fig. 6,d) the cooling jets cling to the blade suction side whereas in the case of the mesh of the group 4 (Fig. 6,e) they penetrate much further into the main flow.

Table 4 shows the kinetic energy losses in the cascade for the different meshes. In the case of the computations without taking into account film cooling, there is a tendency of convergence when mesh is refined. The absolute error in determining the kinetic energy losses in the case of the mesh of the group 3 compared with the case of the mesh of the group 4 is about 0.001 (0.1 percentage points). At the same time, full convergence appears to be still not reached in the case of the computations taking into account film cooling. For the coarser mesh, where the cooling jets are located within the suction side boundary layer, the losses are close to those in the flow without film cooling. For the finer mesh, the cascade losses increase appreciably that can be explained by a significant penetration of the cooling jets into the main flow.

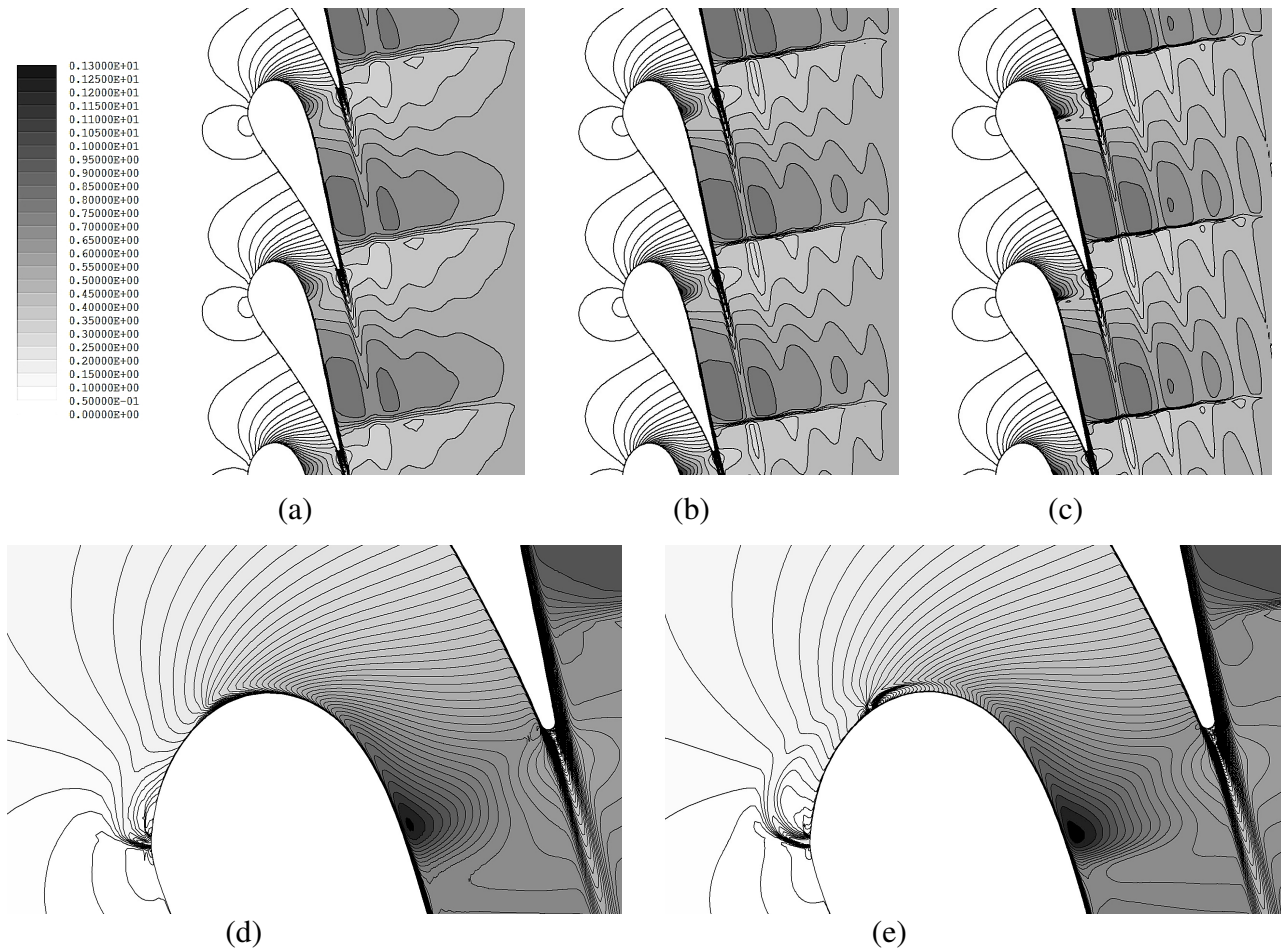

*Fig. 6. The Mach number contours at the midspan section of the TFAST turbine cascade.*
(a), (b), (c) without film cooling ($6.8 \cdot 10^5$, $5.4 \cdot 10^6$ and $4.35 \cdot 10^7$ cells);
(d), (e) with film cooling ($5.4 \cdot 10^6$ and $4.35 \cdot 10^7$ cells)

Table 4

The kinetic energy losses in the TFAST turbine cascade

| Mesh group | 1 | 2 | 3 | 4 |
|---|---|---|---|---|
| Number of cells | $8.5 \cdot 10^4$ | $6.8 \cdot 10^5$ | $5.4 \cdot 10^6$ | $4.3 \cdot 10^7$ |
| w/o film cooling | 0.0653 | 0.0575 | 0.0524 | 0.0512 |
| with film cooling | – | – | 0.0567 | 0.0577 |

## 7. Computational time

Table 5 gives the computational time of all considered flow cases and some additional data related therewith. Also some comments concerning the special conditions of the computational experiments are presented below. The data on the computational time per one time step and number of time steps are given for the finest mesh of each considered flow case (the highest nesting level of SMA).

Table 5

Computational time

| Cascade/stage | VKI-Genoa | ABB-Saturn | TFAST turbine | TFAST compressor |
|---|---|---|---|---|
| Mesh | $4.2 \cdot 10^6$ | $3.07 \cdot 10^7$ | $4.35 \cdot 10^7$ | $2.48 \cdot 10^8$ |
| Dimensions | 128×128×256 | 2×194×208×384 | 368×224×528 | 496×496×1008 |
| Number of mesh levels | 4 | 5 | 5 | 5 |
| Memory, GB | 0.75 | 2×2.8 = 5.6 | 9.5 | 3×16.4 = 49.2 |
| Time per one time step, s | 15 | 59 | 157 | 121 |
| Number of time steps | 60000 | 80000 | 100000 | 200000 |
| Time, days | 12 | 60 | 195 | 370 |

The parallelization of the computations of flow through the ABB-Saturn turbine stage was carried out using the standard tools of the solver F (one blade row per one CPU core). Contrariwise, we performed the computations of flow within the TFAST compressor cascade (which has the straight blade and flat plate endwalls) using the truncated computational domain of one half of the blade span and the symmetry conditions at the cutting line. When parallelizing the computations in this flow case, the computational domain was divided in the axial direction into three blocks. Since the communication between the blocks is time-consuming, the data exchanges were carried out once in the specified number of time steps. The ENO derivatives and the viscous terms also were "frozen" during several time steps to accelerate the calculations. These features are not implemented yet in the standard version of the code, but it will be included in the future.

We have performed the turbomachinery flow computations using the following PCs operating under OS Windows 7:
- Intel Core i7-4820, 3.7 GHz, RAM 64 GB (the meshes of the groups 4 and 5);
- Intel Core i7-3770, 3.5 GHz, RAM 32 GB or 16 GB (the meshes of the groups 1-4).

## 8. Conclusions

The present study confirms the well-known fact that the mesh scales should match the flow scales, namely the characteristic size of the flow regions with significant gradients of thermodynamic, kinematic and turbulent parameters. On the other hand, the obtained results show that the mesh convergence of the kinetic energy losses requires sufficiently fine meshes of $10^7$ cells and more per one blade channel when using the second order-accurate numerical scheme. A good resolution of shock waves, separation zones, wakes, and tangential discontinuities needs the same meshes. An additional mesh refinement may be necessary due to various small-scale features of flow or flowpath geometry, such as film cooling holes, vortex generators, etc. It should be emphasized again that all aforesaid concerns to the numerical solutions of the RANS equations and more sophisticated turbulent flow models, such as DNS and LES, require further special researches.

Based on the present study, we can conclude the following. Scientific researches of the fine flow patterns require a high accuracy and a detail resolution, so, in this case, preference should be given to fine or very fine meshes of the groups 4 and 5. Since such computations is very time-consuming, it may be considered acceptable to use intermediate meshes of the group 3, if the mentioned above requirement on the near-wall cell size, the number of cells across boundary layer, and the mesh expansion ratio in the wall-normal direction are satisfied. In the case of high-volume industrial computations, the use of intermediate (of the group 3) or even coarser meshes sometimes may be sufficient. However, it should be remembered that such computations often result in the flowpath efficiency increase by only 0.001-0.002 (0.1-0.2 percentage points), which is comparable with or even less than discretization errors. Therefore, the final results of such computations should be always verified using finer meshes.

**Acknowledgement**

The authors are grateful to Prof. Olexander Prykhodko for his interest in this study and useful discussions.